\documentclass[aps,prb,lengthcheck,showpacs,superscriptaddress]{revtex4-1}
\usepackage{graphicx}
\usepackage{amsmath}

\begin{document}

\title{Origin of second-harmonic generation enhancement in optical split-ring resonators}

\author{Cristian Cirac\`i}
\email{cristian.ciraci@duke.edu}
\affiliation{Center for Metamaterials and Integrated Plasmonics, Duke University, Durham NC 27708, U.S.}

\author{Ekaterina Poutrina}
\affiliation{Center for Metamaterials and Integrated Plasmonics, Duke University, Durham NC 27708, U.S.}

\author{Michael Scalora}
\affiliation{C.M. Bowden Research Facility, US Army, RDECOM, Redstone Arsenal, AL 35803, U.S.}

\author{David R. Smith}
\affiliation{Center for Metamaterials and Integrated Plasmonics, Duke University, Durham NC 27708, U.S.}

\date{\today}

\begin{abstract}
We present a study of the second-order nonlinear optical properties of metal-based metamaterials.
A hydrodynamic model for electronic response is used, in which nonlinear surface contributions  are expressed in terms of the bulk polarization.
The model is in good agreement with published experimental results, and clarifies the mechanisms contributing to the nonlinear response.
In particular, we show that the reported enhancement of second-harmonic in split-ring resonator based media is driven by the electric rather than the magnetic properties of the structure.
\end{abstract}


\maketitle

Metamaterials (MMs) are artificially structured media whose collective electromagnetic properties derive from the geometry of sub-wavelength inclusions. To date, the most common MM designs have made use of inclusions formed by conducting materials that function as sub-wavelength electrical circuits. These conductor based MMs have proven adept at mimicking a wide variety of linear electromagnetic responses, providing a new venue to explore otherwise inaccessible concepts \cite{Pendry_Cloak}.
In the context of nonlinear response, however, artificial materials may offer even greater opportunities, due to the inherently inhomogeneous local field distribution that exists within and around MM inclusions. By carefully structuring the inclusion geometry, extremely large field enhancement regions can be produced that can dominate and enhance the effective nonlinear response of the composite.

The enhancement of nonlinear processes by MMs has been demonstrated at radio and microwave frequencies, using packaged components, such as varactor diodes, to introduce nonlinearity into the gaps of metal MM inclusions\cite{Larouche:2010fp}.
However, to achieve nonlinear optical materials at higher wavelengths, a simple scaling of these prototype structures to higher frequencies (e.g., beyond a few terahertz) will not suffice.
First, the response of most metals changes from conductor-like to dielectric-like at frequencies above a few terahertz, with absorption increasing significantly as the fields are able to penetrate further into the metal. Second, packaged semiconductor components are not readily available at frequencies above 100 GHz.

While metals and conductors may possess undesirable properties at optical wavelengths, such as increased absorption, they also possess unique and potentially advantageous properties.
In addition to large field enhancements, metal nanostructures also support intrinsic nonlinearities that relate to the dynamics of free and bound charge carriers.
As a result, metals possess some of the largest nonlinear susceptibilities known.
Examples include the large $\chi^{(3)}$ values of gold or silver, for example, suggesting that metals can serve both to form the linear MM response by tailoring the structure, while serving as the source of nonlinearity for nonlinear optical MMs.

The second-order nonlinearity in metals arises from both volume and surface contributions. Nonlinear surface contributions are strictly related to the response of the electrons within the Thomas-Fermi screening length ($\lambda_{TF}\sim0.1$~nm for gold) from the metal boundaries.
In this sub-nanometer realm, electron-electron interactions become appreciable and non-local effects must be taken into account. 
Moreover, since metals are centrosymmetric, they do not possess an inherent $\chi^{(2)}$ nonlinearity. However, the surface of a metal can break spatial symmetry and provide a mechanism for an effective $\chi^{(2)}$ nonlinearity. This homogenized $\chi^{(2)}$ nonlinear response becomes highly dependent on the metal geometry, making it inherently a MM construct.

A steady stream of works concerning harmonic generation from metallic nano-structures has been published recently \cite{vanNieuwstadt:2006dr,Vincenti:2011wn,Ko:2011ch,Zeng:2009hd,Scalora:2010vn}. Second-harmonic generation (SHG) has also been observed experimentally from  a variety of metal nanoparticle systems \cite{Neacsu:2005cq,Shan:2006jx,Butet:2010es,Canfield:2004vi,Klein:2006tj,McMahon:2006iy,Canfield:2007kz,Feth:2008wo,Kim:2008cb,Valev:2010ck,Utikal:2011jr}, and specifically from optical split-ring resonators (SRRs), which at first appeared to exhibit an anomalously high conversion efficiency\cite{Klein,Klein:2006tj,Niesler:2011hd}.
The magnetic resonances associate with SRRs have raised speculation that the enhanced SHG results from a strong nonlinear magnetic response associated with the Lorenz force\cite{Klein}. Yet, a convincing explanation of the nature of this conversion enhancement remains lacking in the literature.

In this Letter, we show that the basic characteristics of nonlinear optical SRRs may be explained solely by the electric properties of the structure rather than its magnetic response, in contrast to the conclusions drawn from previous works\cite{Klein,Klein:2006tj}. The nonlinear optical response of the charge carriers in the metal here is described by a hydrodynamic model, which includes the effects of pressure associated with the electron gas.
To facilitate the numerical models and remove ambiguities associated with additional boundary conditions, we develop an expression for the effective second-harmonic surface currents in terms of the polarization vector in the bulk regions. This expression allows us to easily study the SHG process from metal particles of arbitrary shape.

In the context of the hydrodynamic model, the electron fluid density, $n({\bf r},t)$, and the electron velocity field, ${\bf v}({\bf r},t)$, satisfy Euler's equation:
\begin{equation}
 \label{euler}
\frac{{\partial {\bf{v}}}}{{\partial t}} + \left( {{\bf{v}}\cdot\nabla } \right){\bf{v}} + \gamma {\bf{v}} = \frac{e}{{m_e^* }}\left( {{\bf{E}} +{\bf{v}} \times {\bf{B}}} \right) - \frac{{\beta ^2 }}{n}\nabla n,
\end{equation}
along with the continuity equation, $\nabla \cdot {\bf J}=-e{\dot n}$, with ${\bf J}=en{\bf v}$.
In Eq.~(\ref{euler}), $m_e^*$ is the effective electron mass, and $\gamma$ is the electron collision rate.
The last term in the equation is due to the electron gas pressure, here described within the Thomas-Fermi model, with $\beta$ proportional to the Fermi velocity $v_F$ .
Combining  the continuity equation with Eq.~(\ref{euler}), and expanding all fields in a perturbative manner one finds that the free electron polarization ${\bf  \dot P}={\bf J}$ satisfies the following set of inhomogeneous equations\cite{Scalora:2010vn}:
\begin{subequations}
 \label{free_el}
\begin{equation}
 \label{free_el_1}
\beta ^2 \nabla \left( {\nabla \cdot{\bf{P}}_1 } \right) + \left( {\omega ^2  + i\omega \gamma } \right){\bf{P}}_1  =  - \frac{{n_0 e^2 }}{{m_e^* }}{\bf{E}}_1,
 \end{equation}
\begin{equation}
\label{free_el_2}
 \beta ^2 \nabla \left( {\nabla \cdot{\bf{P}}_2 } \right) + \left( {4\omega ^2  + 2i\omega \gamma } \right){\bf{P}}_2  =  - \frac{{n_0 e^2 }}{{m_e^* }}{\bf{E}}_2  +  {\bf S}_{\rm NL} ,
 \end{equation}
 \end{subequations}
where the nonlinear source, ${\bf S}_{\rm NL}$, is given by:
\begin{equation}
 \label{S_NL}
\begin{array}{l l}
 {\bf S}_{\rm NL}  = & \frac{e}{{m_e^* }}{\bf{E}}_1 \left( {\nabla \cdot{\bf{P}}_1 } \right) + \frac{{i\omega e}}{{m_e^* }}{\bf{P}}_1  \times {\bf{B}}_1  \\ 
 &  - \frac{{\omega ^2 }}{{n_0 e}}\left[ {\left( {\nabla \cdot{\bf{P}}_1 } \right){\bf{P}}_1  + \left( {{\bf{P}}_1 \cdot\nabla } \right){\bf{P}}_1 } \right] . 
\end{array}
 \end{equation}
The subscripts refer to fundamental and second-harmonic fields respectively.
In deriving Eqs.~(\ref{free_el}) a harmonic time dependence ($e^{-i\omega t}$) has been assumed.
Together with Maxwell's equations, these equations describe the fundamental and second-harmonic polarization vectors, respectively, and hold under the assumption that the fundamental field remains undepleted.
The nonlinear source given by Eq.~(\ref{S_NL}) groups surface and bulk second-harmonic contributions.
Specifically we have the nonlinear Coulomb term, (referred to as a quadrupole-like term by virtue of its form) proportional to ${\bf{E}_1}\left( {\nabla  \cdot {\bf{P}}_1 } \right)$, the magnetic Lorentz force contribution, ${\bf{ P}}_1 \times {\bf{B} _1}$, and convective terms  $ ( {\nabla  \cdot {\bf{ P}}_1 } ){\bf{ P}}_1$ and $( {{\bf{ P}}_1  \cdot \nabla } ){\bf{ P}}_1 $.

The effect of the electron gas pressure on the polarization vectors ${\bf{ P}}_1$ and ${\bf{ P}}_2$ in Eqs.~(\ref{free_el}) is a linear, nonlocal contribution of the form $ \beta^2\nabla \left( {\nabla \cdot{\bf{P}} } \right)$.
This term has been predicted to be responsible for unusual, resonant-like phenomena above the plasma frequency\cite{Ruppin:2001ba,Raza:2011ioa}.
In general, this term becomes important in a region of order of $\lambda_{TF}$ in the vicinity of the metal surface, where electron-electron interactions dominate the nonlinear surface sources given by Eq.~(\ref{S_NL}).
However, the linear non-local term may be safely neglected in Eq.~(\ref{free_el_2}), since it does not affect the amount of generated harmonic\cite{Sipe:1980vz}.

The presence of spatial derivatives (nonlocality) in the description of  the fundamental polarization vector of Eq.~(\ref{free_el_1}) requires the specifications of additional boundary conditions to solve the electromagnetic boundary value problem\cite{Halevi:1984va,Maystre:1986wi}.
From the continuity equation and Gauss' theorem one obtains that ${\bf{\hat n}} \cdot \bf{P}=0$ at the boundary\cite{Jewsbury:1981uu,Raza:2011ioa}.
A consequence of the linear pressure term is the non-zero extension of the induced surface charge inside the metal.
That is, the induced electron charge density, $\frac{1}{e}\nabla \cdot {\bf P}_1$ , is zero in the bulk region, and rapidly changes near the metal surface, where it  reaches its maximum value.
This behavior reflects the variation experienced by the normal component of ${\bf P}_1$, which continuously goes to zero at the metal boundary.

The hydrodynamic model is a simplistic model of electrons that nevertheless gives a fairly accurate description of linear and nonlinear processes occurring at the surface of metallic structures.
However, the simultaneous manifestation of microscopic and macroscopic scales in the problem makes the resolution of the system of (nonlocal and nonlinear) equations quite complex, and ordinarily necessitates considerable computational resources even for particles whose dimensions are a few tens of nanometers.
Deep inside the metal, in the bulk region, the electron pressure may be neglected, since it leads to corrections of order $(\lambda_{TF}/\lambda)^2 \ll 1$ \cite{Sipe:1980vz}.
On the other hand, second-harmonic conversion efficiency strongly depends on the behavior of ${\bf P}_1$ at the metal surface where the impact of the pressure term may be critical.
Moreover, if the electron pressure in Eq.~(\ref{free_el_1}) is neglected at the surface, the theory becomes inherently ambiguous, as pointed out by Sipe \textit{et al.}\cite{Sipe:1980vz}.

To understand where the ambiguity arises consider the nonlinear Coulomb term in Eq.~(\ref{S_NL}), which is proportional to $(\nabla \cdot  {\bf P}_1 ){\bf E_1}$.
In the free electron limit ($\beta=0$) the normal component of the electric field $E_n=\hat{{\bf n}}\cdot{\bf E}_1$ changes discontinuously across the metal boundaries.
This discontinuity is attributed to charge accumulation occurring at the surface, which assumes the form of a Dirac delta function, namely, $\nabla \cdot {\bf P}_1 =P_n \delta_0(\hat{{\bf n}}\cdot{\bf r})$, where $P_n$ is the the bulk polarization component normal to the surface.
It follows that the quantity $(\nabla \cdot {\bf P}_1)E_n=P_n\delta_0E_n$ is not well defined and cannot be integrated.
The same analysis applies to the convective terms.
In early works this problem was circumvented by introducing phenomenological coefficients of order unity\cite{Rudnick:1971wb}, or through an effective plasma frequency\cite{Sipe:1980vz}, that incorporates the details of the charge distribution near the surface, an effect that was neglected in Ref.~[\citenum{Zeng:2009hd}].
However, an analysis of the effect of pressure on the amount of converted energy shows that the total SHG converges to an asymptotic value as $\beta$ tends toward zero.
We verified that for typical values of $\beta$ (of order of $10^6$~m/s ), the actual converted energy differs by not more than 3\% from its asymptotic value.
Exploring the limit for $\beta \to 0$ seems then a very good way to get an approximate solution without having to solve the complex nonlocal equations, as discussed below.

In this limit the surface layer is so small compared to the size of the nanoparticle that only the derivative along the direction,  $\xi={\bf r}\cdot {\bf \hat{n}}$, normal to the metal boundary matters.
Let us choose the metal boundary such that for $\xi \le 0$ the electron pressure may be neglected ($\beta=0$); the region $0<\xi\le l$, where $l \sim \lambda_{TF}$, represents the surface layer where the pressure is important.
In this region, the parallel component of the vector ${\bf{P}}_1$ is nearly constant,
$P_1^\parallel (\xi) \simeq P_1^\parallel(0^-)$, while its normal component $P_1^\bot$ may be written as:
\begin{equation}
 \label{P_guess}
P_1^\bot (\xi ) = P_1^\bot (0^-) \sigma (\xi ) ,
  \end{equation}
where $\sigma (\xi )$ is an unknown, rapidly varying function that goes from $\sigma(0)=1$ to $\sigma(l)=0$ (as imposed by boundary conditions).
It is straightforward to show that starting with Eq.~(\ref{free_el_2}), that the nonlinear surface polarization is given by:
\begin{equation}
 \label{pol_s}
{\bf{P}}_2^S  =  - \frac{1}{{2n_0 e}}\left[ {\left( {\nabla \cdot{\bf{P}}_1 } \right){\bf{P}}_1  + \frac{\omega }{{2\omega  + i\gamma }}\left( {{\bf{P}}_1 \cdot\nabla } \right){\bf{P}}_1 } \right],
 \end{equation}
where the second-harmonic pressure term, $\beta ^2 \nabla( {\nabla \cdot{\bf{P}}_2 } )$, has been neglected, as already pointed out. 
In writing Eq.~(\ref{pol_s}) we have used Eq.~(\ref{free_el_1}) to express the electric field ${\bf E}_1$ as a function of ${\bf P}_1$.
The crucial quantity here is the effective nonlinear surface current density, defined as: ${\bf{K}}_{NL} \equiv 2i\omega  \int_{\xi=0}^{\xi=l} {{\bf{P}}_2^{S} ({\bf r})d{\bf r}}$, where the integral is performed across the surface layer.
Using Eq.~(\ref{pol_s}) we obtain:
\begin{equation}
 \label{J_NL}
{\bf{K}}_{NL}  = \frac{i\omega}{n_0 e}\left[ {{\bf{\hat t}}\left( {P_1^ \bot  P_1^\parallel  } \right) + {\bf{\hat n}}\frac{1}{2}\frac{3\omega  + i\gamma }{2\omega  + i\gamma} (P_1^ \bot)^2  }\right],
\end{equation}
where $\int_0^l {\sigma 'd\xi }=-1$ and  $\int_0^l {\sigma \sigma 'd\xi }  =-1/2$ have been used, with the prime representing the derivative with respect to $\xi$, and the unit vector ${\bf{\hat t}}$ pointing in the direction ${\bf{\hat n}}\times{\bf{P}}_2^S$.
Eq.~(\ref{J_NL}) states that surface contributions to the second-harmonic polarization may be approximated by an effective nonlinear current sheet at the surface of the nanoparticle.
These currents are related to the polarization values in the bulk region and do not require the resolution of nonlocal equations.
That is, the volume sources are calculated by assuming $\beta=0$ while the surface nonlinear currents are given by our Eq.~(\ref{J_NL}) .
In the limit of validity, this approach provides a description of the SHG process that may be easily implemented in full 3D simulations.
We tested this approach in the case of SHG from a silver slab and compared our results with experimental data\cite{ODonnell:2005hc}.
The results obtained using Eq.~(\ref{J_NL}) agreed very well with both the solution given by the complete Eqs.~(\ref{free_el}) and the experimental data.

\begin{figure}
\includegraphics[width=8.6cm]{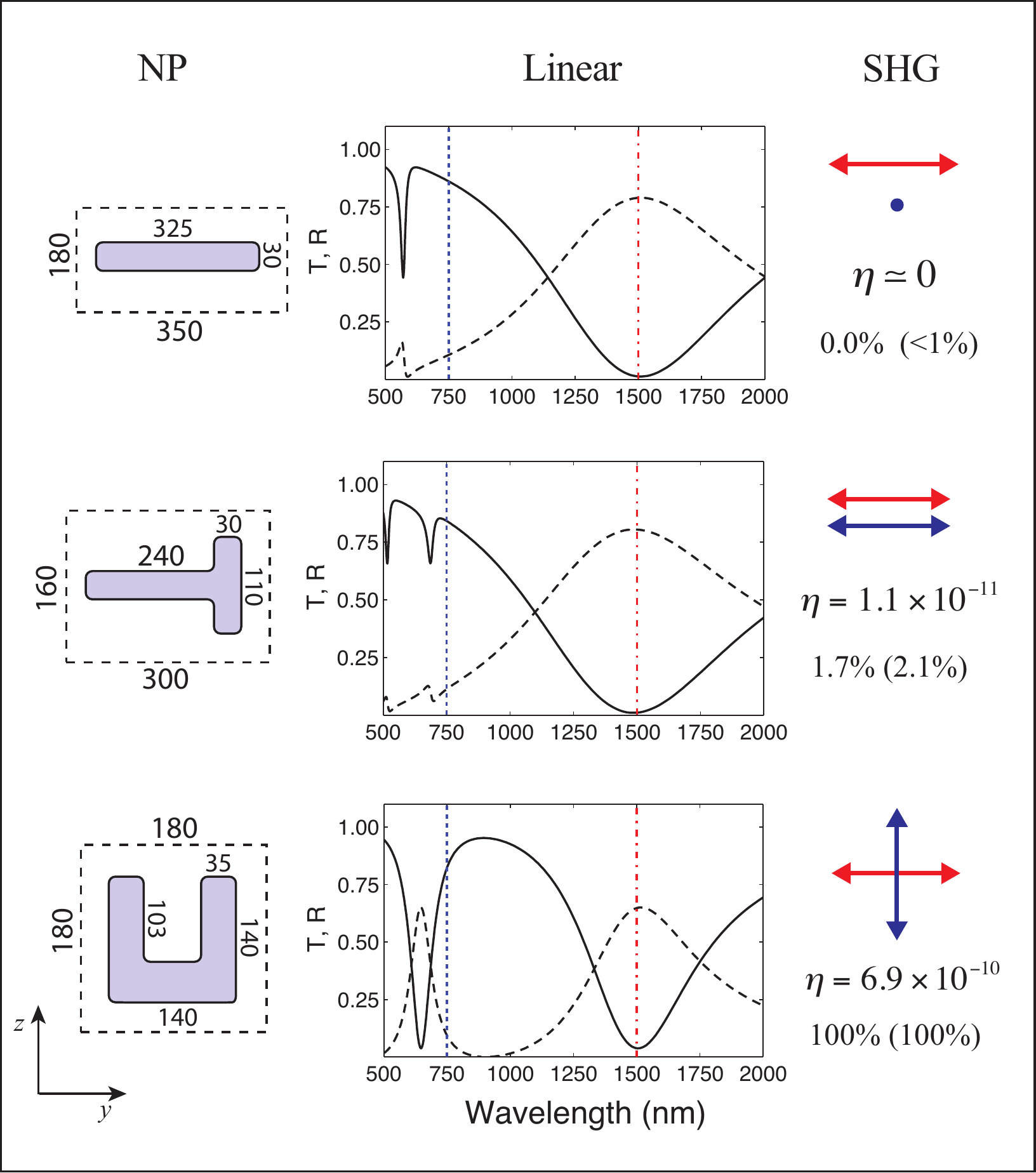}          
\caption{(Color on line) Second-harmonic conversion efficiency for different 3D gold nanoparticles.
All the particles are 20~nm thick.
The linear transmittance (solid line) and reflectance (dashed line) at normal incidence are shown in the second column for the variety of nanoparticles.
The vertical lines indicate the fundamental (dotted red line), and the second-harmonic (blue dashed-dot line) wavelength, respectively.
In the third column, the arrows denote the polarization state. The fundamental field is represented in red (light gray), the second-harmonic field is blue (gray). 
The relative conversion efficiencies normalized with respect to the U-shaped nanoparticle efficiency are also reported (the corresponding experiment of Ref.~\cite{Klein:2006tj} are shown inside brackets). 
The following values for the parameters have been used: $m_e^*=m_e$, $n_0=5.7\times10^{22}$~cm$^{-3}$, and $\gamma=1.07\times 10^{14}$~s$^{-1}$.}
\label{fig1}
\end{figure}

3D metal nanoparticle systems are of considerable interest for nonlinear media, because both the field enhancement regions and the surface morphology strongly impact SHG. In one of the first experiments of its kind, planar arrays of varying MM inclusions were shown to produce second-harmonic light, with efficiencies that varied according to the shape of the nanostructured inclusions\cite{Klein:2006tj}.
To investigate the mechanism of SHG from metal nanoparticles, we choose the same types of nanoparticles studied in Ref.~[\citenum{Klein:2006tj}].
Our results are summarized in Fig.~\ref{fig1} for the different kinds of nanoparticles. To ease the computational burden and to simplify the system studied, we assumed the nanoparticles to be surrounded by air rather than including the substrate material used in the experiments.
We do not expect significant changes in the observed mechanism of SHG due to the exclusion of the substrate material. The structure studied extends periodically in the $x$ and $y$ directions, so that only a single unit cell is needed in the computational space.
To avoid possible numerical artifacts due to the field localization near metal corners, we considered rounded corner geometries with a radius of curvature of 5~nm.
The geometrical parameters were chosen so that the nanoparticles would display a resonance around $\lambda_{\rm FF}=1.5~\mu$m, to which the fundamental field is tuned.
The conversion efficiencies, $\eta$, assume an average pump intensity of $\sim55$~MW/cm$^2$ with the electric pumping field polarized along the $x$-direction.

We find qualitatively good agreement with the experimental data of Ref.~[\citenum{Klein:2006tj}], where second- and third-harmonic generation were experimentally investigated for rectangular, T- and U-shaped gold nanoparticles respectively.
We find the relative efficiencies for normal incidence normalized with respect to the U-shaped nanoparticle efficiency to be about 1.7\% for T-shaped nanoparticles (compared with 2.1\% for the experimental data), while the rectangular nanoparticle showed a $\sim0$ conversion efficiency, as one might expect due to the absence of symmetry breaking. 

\begin{figure}
\includegraphics[width=8.2cm]{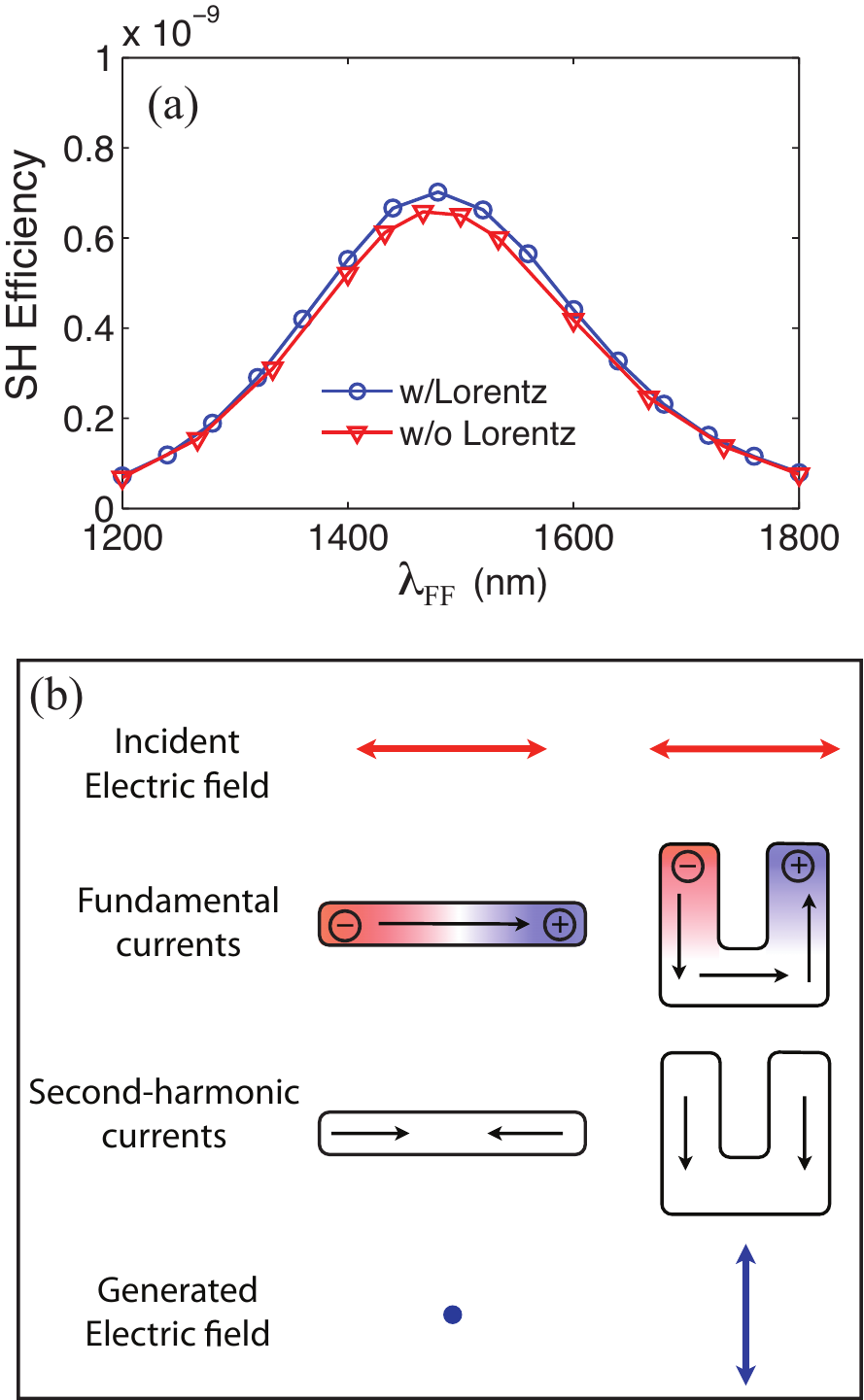}          
\caption{(Color on line) (a) Comparison of the second-harmonic conversion efficiency with and without Lorentz force contributions for the U-shaped nanoparticle of Fig.~\ref{fig1}.
(b) Sketch of the fundamental and second-harmonic currents flowing in the rectangular and U-shaped nanoparticles respectively.
For the symmetric nanoparticle  the  nonlinear currents cancel out and no second-harmonic field is produced.
Instead, for the U-shaped nanoparticle the currents oscillate in phase and radiate to the far-field. }
\label{fig2}
\end{figure}

As in Ref.~\cite{Klein:2006tj} we find that an array of U-shaped gold nanoparticles can enhance the SHG efficiency by about two orders of magnitude with respect to other non-centrosymmetric nanoparticles, such as T-shaped particles for example.
Moreover, the polarization state of the generated field is rotated by 90$^\circ$ for a pumping electric field polarized along the $x$-direction. This polarization flip is not surprising, because symmetry breaking occurs along the $y$-direction. However, the nature of the conversion enhancement remains an unsettled point in the literature.
In Ref.~\cite{Klein} the authors speculated that this effect arose as a result of a strong nonlinear magnetic response associated with the Lorenz force. However, the Lorentz force, proportional to ${\bf{P}_1} \times {\bf{B} _1}$, is but one of several influential contributions, Eq.~(\ref{S_NL}). Within the context of our model, by simply switching off the Lorentz force contribution to the generated field we find that the conversion efficiency is only marginally affected.
The effect is shown in Fig.~\ref{fig2}a, where the second-harmonic conversion efficiency is calculated as a function of the pumping wavelength, $\lambda_{\rm FF}$, with and without the Lorentz force term.
Therefore, the magnetic response of the U-shaped nanoparticle does not seem to play an important role in the description of the SHG process, at least assuming the hydrodynamic model. 

The origin of the conversion enhancement may be explained by studying the nonlinear electrical properties of the nanoparticle. 
If the Lorentz contribution is neglected, all the electric nonlinear contributions are basically proportional to the product of the electric field ${\bf E}$ and the induced charge distribution $\rho=-en=\nabla \cdot {\bf P}$, as may easily be deduced from Eq.~(\ref{free_el_2}).
In Fig.~\ref{fig2}b two parallel situations are depicted: i) the rectangular nanoparticle and ii) the U-shaped nanoparticle.
The arrows indicate the electric field at the surface, while the color map represents the charge distribution.
At the fundamental frequency both structures behave similarly.
The external electric field of the electromagnetic radiation produces a gradient in the electrical potential that causes charges to migrate from one pole to the other. To a first approximation, the resulting second-harmonic polarization is proportional to the product between the charge distribution and the electric field, that is: ${\bf P}(2\omega)\propto \rho(\omega) {\bf E}(\omega)$.
As shown in the sketch of Fig.~\ref{fig2}b, the nonlinear polarization will result in a second-harmonic electric field that cancels out for the centrosymmetric nanoparticle, resulting in a near-zero second-harmonic efficiency. On the other hand, for the U-shaped nanoparticle the resulting harmonic fields oscillate in phase and the produced field can radiate to the far-field. These findings lead us to suggest that the characteristic emission of nonlinear optical SRRs is driven by the electric properties of the structure rather than its magnetic response, as originally conjectured.

In conclusion, we have discussed the influence of the electron gas pressure on SHG and find a new way to express nonlinear surface contributions in terms of the polarization vector in bulk regions.
Using this approach we have analyzed the SHG in several types of 3D gold nanoparticles and demonstrated that the origin of the previously experimentally reported enhanced SHG from SRR-based nano-structures is mostly electrically driven. 
This development should help simplify the investigation of the SHG process in full 3D metal structures, and enable the investigation of arbitrary nano-structured plasmonic media, offering a viable design approach to integrated, efficient nonlinear optical media.

\begin{acknowledgments}
The authors would like to thank Yaroslav Urzhumov for valuable discussions. This work was supported by the Air Force Office of Scientific Research (AFOSR, Grant no. FA9550-09-1-0562)
\end{acknowledgments}

\end{document}